\documentclass{jfm}

\usepackage{amsmath,amsfonts} 
\usepackage{graphicx,wasysym}
\usepackage{multirow}
\usepackage{color}

\newcommand{\fr}[1]{\ref{fig:#1}}
\newcommand{\er}[1]{(\ref{eq:#1})}

\newcommand{\sr}[1]{\ref{sec:#1}}
\newcommand{\pd}[2]{\frac{\partial #1}{\partial #2}}
\newcommand{\pdi}[2]{\partial_{#2} {#1}}

\newcommand{\eq}[1]{\begin{align} #1 \end{align}}

\begin{document}

\newtheorem{lemma}{Lemma}
\newtheorem{corollary}{Corollary}

\shorttitle{On the spreading of impacting drops} 
\shortauthor{S. Wildeman et al} 

\title{On the spreading of impacting drops}

\author
 {
 Sander Wildeman\aff{1}
  \corresp{\email{swildeman@gmail.com}},
  Claas Willem Visser\aff{1},
  Chao Sun\aff{1,3}
  \and 
  Detlef Lohse\aff{1,2}
  }

\affiliation
{
\aff{1}
Physics of Fluids group, MESA+ Institute and J. M. Burgers Centre for Fluid Dynamics, University of Twente, 7500 AE Enschede, The Netherlands
\aff{2}
Max Planck Institute for Dynamics and Self-Organization, 37077, G\"ottingen, Germany
\aff{3}
Center for Combustion Energy and Department of Thermal Engineering, Tsinghua University, 100084 Beijing, China
}

\maketitle

\begin{abstract}
The energy budget and dissipation mechanisms during droplet impact on solid surfaces are studied numerically and theoretically. We find that for high impact velocities and negligible surface friction at the solid surface (i.e. free-slip), about one half of the initial kinetic energy is transformed into surface energy, independent of the impact parameters and the detailed energy loss mechanism(s). We argue that this seemingly universal rule is related to the deformation mode of the droplet and is reminiscent of pipe flow undergoing a sudden expansion, for which the head loss can be calculated by multiplying the kinetic energy of the incoming flow by a geometrical factor. For impacts on a no-slip surface also dissipation in the shear boundary layer at the solid surface is important. In this case the geometric head loss acts as a lower bound on the total dissipation (i.e. the spreading on a no-slip surface approaches that on a free-slip surface when the droplet viscosity is send to zero).

This new view on the impact problem allows for simple analytical estimates of the maximum spreading diameter of impacting drops as a function of the impact parameters and the properties of the solid surface. It bridges the gap between previous momentum balance approaches and energy balance approaches, which hitherto did not give consistent predictions in the low viscosity limit. Good agreement is found between our models and experiments, both for impacts on ``slippery'' or lubricated surfaces (e.g. Leidenfrost droplet impacts and head-on droplet-droplet collisions) and for impacts on no-slip surfaces.
\end{abstract}

\section{Introduction}

The common event of a liquid drop colliding with a solid surface is a very rich phenomenon: droplets can splash, spread or bounce \citep{Rein1993, Yarin2006, Marengo2011, Quere2013, Josserand2016}, often entrapping air in the process \citep{Chandra1991,Thoroddsen2005}. But despite the great experimental progress on the droplet impact problem \citep{Thoroddsen2008} consensus has yet to be reached on many of its basic laws. One of the very basic questions (see figure~\fr{problem}) is to find out the maximum spreading diameter $D_m$ of the droplet, given its initial diameter $D_0$, its impact velocity $V_0$, the properties of the liquid (density $\rho$, surface tension $\gamma$, and viscosity $\mu$), and the conditions at the solid surface: the slip length $\lambda$ \citep{Lastakowski2014} and the (dynamic) liquid-solid contact angle~$\theta$ \citep{Sikalo2005,Antonini2012}. The influence of the surrounding air is often negligible in the overall spreading dynamics \citep{Visser2015}. Using dimensional analysis, the problem can be reduced to finding the spreading ratio $\tilde D_m \equiv D_m/D_0 = f(\text{We}, \text{Re}, \lambda/D_0, \theta)$, where
\eq{
\text{We} \equiv \frac{\rho D_0 V_0^2}{\gamma}\text{\;\;\;and\;\;\;} \text{Re} \equiv \frac{\rho D_0 V_0}{\mu}\nonumber
}
are the impact Weber and Reynolds numbers, which reflect,  respectively, the importance of the surface tension and viscosity with respect to the initial inertia of the drop.

Knowing the maximum surface coverage that an impacting drop can reach is of great importance in a wide variety of applications:  from ink-jet printing \citep{VanDam2004}, spray cooling \citep{Kim2007} and self-cleaning surfaces \citep{Blossey2003}, to blood spatter analysis in crime scenes \citep{Laan2014} and anti-icing of aircraft wings \citep{Mishchenko2010}. All these applications deal with high inertia impacts ($\text{We} \gg 1$, $\text{Re} \gg 1$) in which the droplets greatly deform. This regime will be the focus of this work.

\begin{figure}
\centerline{\includegraphics{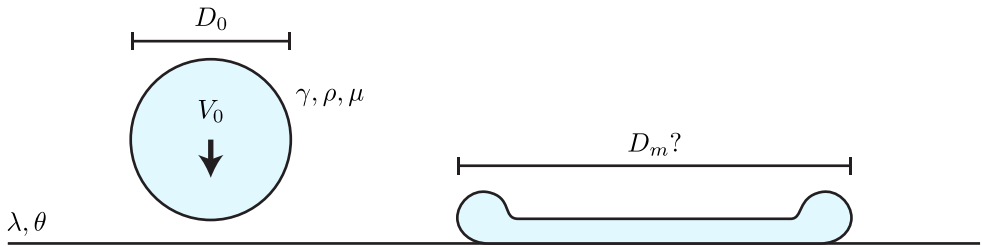}}
\caption{The canonical problem in droplet impact: A droplet hits a solid surface with initial velocity $V_0$ and diameter $D_0$. What will be its maximum spreading diameter $D_m$ as a function of the liquid properties (surface tension $\gamma$, viscosity $\mu$, density $\rho$) and solid properties (slip length $\lambda$, (dynamic) contact angle $\theta$)? \label{fig:problem}}
\end{figure}

In the literature one can identify two strategies to estimate $\tilde D_m$: (i) an energy balance or (ii) a force/momentum balance. In the first case it is usually assumed that the initial kinetic energy is mainly consumed by the creation of new surface area and by viscous dissipation in the shear boundary layer growing into the droplet from the solid surface \citep{Chandra1991,Pasandideh-Fard1996,Aziz2000, Ukiwe2005, Lee2016}. Although this approach works reasonably well in the viscous limit, where the boundary layer dominates the dynamics, it has been found that it severely overestimates the maximum spreading for cases in which viscous dissipation is supposed to be negligible \citep{Chandra1991,Jiang1992,Clanet2004, Attane2007}. Viscosity free models based on a momentum balance, on the other hand, predict the spreading in this regime quite well \citep{Clanet2004,Eggers2010, Lastakowski2014}. The question is whether we can bring the two approaches in line.

As has previously been noted by \citet{Villermaux2011,Roisman2011}, such a discrepancy between energy and momentum approaches also occurs in a closely related fluid dynamics problem: that of the growth of a hole in a punctured liquid film \citep{Taylor1959, Culick1960}. Taylor found that, after a start-up period \citep{Sunderhauf2002, Savva2009}, a thick cylindrical rim forms that consumes the film at a constant retraction velocity $V_c$. Using momentum and mass conservation he derived that for a film of thickness $h$ this velocity must be $V_c = \sqrt{2\gamma/h\rho}$, in good agreement with experiments. One year later, Culick, seemingly unaware of Taylor's result, wrote a short comment on the same problem, finding the same steady state velocity. Moreover, he emphasized that this velocity is lower by a factor of $\sqrt{2}$ than what one would get from an energy balance between the kinetic energy of the rim and the surface energy consumed by it. That is, \emph{one half} of the available surface energy never makes it into kinetic energy. The rest must be dissipated somewhere, somehow. Although viscous stresses do not enter explicitly in the momentum balance for the retraction problem, Culick argued \textit{a posteriori} that they have to be concentrated near the rim entrance, where fluid from the stationary film is accelerated to match the velocity in the rim. The volume over which the associated dissipation occurs must then scale in such a way with viscosity as to make the final answer independent of it. Interestingly, the problem can also be related to the head loss in a pipe flow undergoing a sudden expansion \citep{Villermaux2011}. In this situation the energy is generally lost to recirculation eddies forming downstream of the expansion section. But also here the final dissipation is insensitive to the details and can be directly calculated by multiplying the initial kinetic energy by a geometric factor \citep{Batchelor1967}.

In this work we use direct numerical simulations to demonstrate that a similar universal energy loss occurs for droplets which greatly deform during impact (i.e. at high Weber numbers). Taking this loss into account in a simple energy balance allows us to predict (without fitting parameters) the spreading diameters found in impact situations in which surface friction is negligible (as is the case, for example, for Leidenfrost droplets, for head-on droplet-droplet collisions or for impacts at very high Reynolds numbers). To make the story complete we extend this model to spreading on ``dry'' solid surfaces at intermediate Reynolds numbers, for which the flow is also hindered by the no-slip boundary condition.


\section{Method}

The calculation and visualization of energy dissipation inside an impacting droplet requires detailed information about the flow field. To this end, we used the open source volume of fluid solver Gerris \citep{Popinet2009} to simulate the impact event. Gerris solves the incompressible Navier-Stokes equation on an adaptive grid \citep{Popinet2003} and is renowned for its physically sound treatment of interface dynamics~\citep{Popinet2009}. This makes Gerris well suited for simulating two-phase flows involving a multitude of length scales, such as droplet impact \citep{Thoraval2012, Thoraval2013}.

In each simulation an axisymmetric droplet, surrounded by a gas, was set to collide with an impermeable boundary of the domain. For the high Weber number cases the minimum refinement levels were 10 in the interior of the droplet and 13 at its surface (corresponding to $N = 2^{10}/3 \approx 340$ and $N = 2^{13}/3 \approx 2730$ cells in one droplet diameter). During the simulations, these resolutions were adaptively increased in regions of high vorticity and strong curvature (up to levels 12 and 16, respectively). The liquid-air density and viscosity ratios were set to $\rho/\rho_g = 1000$ and $\mu/\mu_g = 50$, mimicking water in air. In \citet{Visser2015} we compared the droplet profiles obtained with this numerical scheme to experimental recordings with high temporal and spatial resolution. Good agreement was found in all cases.

The numerical approach makes it relatively straightforward to impose different slip conditions and liquid-solid contact angles at the solid surface. For maximal contrast, we studied the limiting cases of no-slip (slip length $\lambda \rightarrow 0$, radial velocity $v_r=0$ at $z=0$) and free-slip boundary conditions ($\lambda \rightarrow \infty$, $\pdi{v_r}{z} = 0$ at $z=0$). In the no-slip simulations the interface of the impacting droplet often did not coalesce with the surface, so that a thin layer of air remained between the surface and the liquid. During the spreading phase the effective slip length introduced by this air layer was always small compared to the thickness of the shear boundary layer in the liquid, so that it did not noticeably change the overall dynamics as compared to true no-slip (we checked this for a few cases by forcing the droplet to coalesce at initialization). This non-coalescence behaviour fixed the contact angle for the no-slip droplets to $\theta = 180^\circ$. For the free-slip case we used both $\theta = 180^\circ$ and $\theta = 90^\circ$. The latter case describes the head-on collision of two equally sized droplets that coalesce upon contact \citep{Roisman2009}.

With the axisymmetry assumed in the simulations we obviously cannot capture non-axisymmetric effects like azimuthal rim destabilization and break-up~\citep{Villermaux2011}. However, this idealization allows us to test basic models in the very high ($> 1000$) Weber number regime, which is difficult to access experimentally.

\section{Deformation modes: from puddle to pizza \label{sec:defmode}}

When the Reynolds number is high, the qualitative mode of deformation of an impacting droplet mainly depends on its Weber number. In figure \fr{regimes} we show numerically obtained profiles and internal flow patterns (at maximal spreading) for four orders of magnitude in Weber number ($\text{We} \in \{0.3, 3, 30, 300\}$). The results for no-slip and free-slip boundary conditions are shown side by side to facilitate a direct comparison. The Reynolds number was $500$ in all eight cases.

\begin{figure}
\centerline{\includegraphics{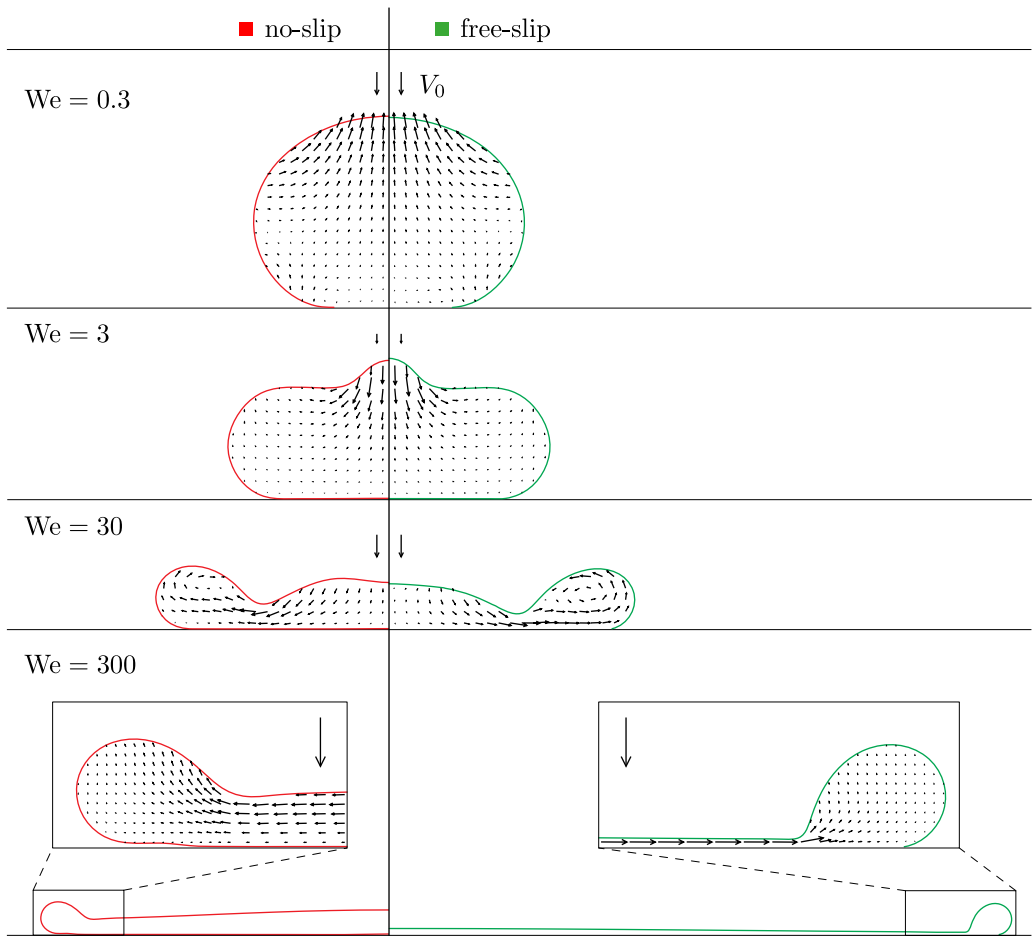}}
\caption{Droplet profiles and flow patterns at maximum spreading obtained from our simulations as a function of the Weber number. Impacts on a no-slip surface are shown on the left and impacts on a free-slip surface on the right. The Reynolds number was 500 for all cases (and $\theta = 180^\circ$). The downward arrows above each droplet depict the initial impact velocity $V_0$. \label{fig:regimes}}
\end{figure}


For $\text{We} = 0.3$ the surface tension of the droplet dominates over its inertia and the droplet therefore deforms only slightly from spherical during impact. The close similarity between left and right profiles in figure \fr{regimes} indicates that there is no large effect of the slip length in this case. This can be easily understood from the fact that, at this low Weber number, the impact time is too short for the shear boundary layer at the no-slip surface to grow appreciable. The thickness of the boundary layer at the time of maximum deformation, $\tau_m$, can be estimated as the distance of momentum diffusion $H_b \approx \sqrt{\nu \tau_m}$ \citep{Roisman2009-2,Eggers2010,Visser2015}, with $\nu \equiv \mu/\rho$ the kinematic viscosity of the liquid. Comparing this to the central height $H_c(\tau_m) \approx D_0$ of the droplet, gives $H_b/D_0 \approx \sqrt{\tilde \tau_m/\text{Re}}$, where we introduced $\tilde \tau_m = (V_0/D_0) \tau_m$. From our simulations we find that $\tilde \tau_m \approx 0.3$ for $\text{We} = 0.3$, so that $H_b/D_0 \approx 0.02 \ll 1$.

As the Weber number is increased beyond $\text{We} = 1$, the droplet gradually starts to lose its spherical shape. At $\text{We} = 3$, the spread-out droplet resembles a puddle with a flattened top and a rounded edge. Although the droplet height is now lower, $H_c(\tau_m) \approx D_0/2$, and the dimensionless spreading time has increased, $\tilde \tau_m \approx 0.5$, there is still no noticeable effect of the no-slip condition. Indeed, a quick calculation gives $H_b/H_c \approx 0.06 \ll 1$. In contrast to the $\text{We} = 0.3$ case, the central part of the $\text{We} = 3$ droplet still moves down at the moment of maximum spreading. Markedly, near the top surface the downwards flow velocity is even higher than the initial impact velocity. When the simulation is continued, this ``jet'' further penetrates into the bulk, creating an indentation in the drop's upper surface before the droplet finally retracts and rebounds. It has been shown that around $\text{We} = 10$ this indentation almost completely perforates the droplet, giving it the appearance of a doughnut \citep{Renardy2003}. Under these special conditions a fast upwards jet is produced as the thick retracting rim closes in on the central hole and squeezes out the air \citep{Bartolo2006}.

Around $\text{We} = 30$ yet another mode of deformation starts to set in. Shortly after impact, a rim (now thin compared to $D_0$) is squeezed out from the bottom part of the droplet, while the rest of the droplet moves on in an almost undisturbed way. As the droplet sinks into the solid, more and more liquid flows out into this rim, which is at the same time pulled back by surface tension. The liquid in the rim whirls around as it is pushed out through the narrow neck connecting the rim and the central part of the droplet \citep{Clanet2004}. As can be seen in figure \fr{regimes}, this vortical motion is still present at the moment of maximum spreading. In this deformation mode we start to see an influence of the slip condition on the spreading. The thickness of the boundary layer $H_b(\tilde \tau_m)/D_0 \approx \sqrt{1.1/500} \approx 0.05$ is not negligible compared to the thickness of the neck, $H_n(\tilde \tau_m)/D_0 \approx 0.2$. As a consequence the neck is slightly thicker in the no-slip case, and the droplet spreads out somewhat less.

When the Weber number is increased by another order of magnitude, to $\text{We} = 300$, the qualitative features of the drop's deformation discussed for $\text{We} = 30$ become more pronounced. In this case a thin lamella bordered by a rim spreads out radially from the impact zone (see also figures \fr{spatial}(a) and (b)). The vortical motions in the rim are now absent (see insets figure \fr{regimes}). Again the bulk of the droplet moves on in an undisturbed way initially, but around $\tilde t \approx 0.5$ it gradually starts to slow down \citep{Eggers2010}. At the moment of maximal extension the droplet has attained the shape of a pizza: a thin, almost flat central part, bordered by a thick cylindrical rim. The maximum diameter and the time in which it is reached now strongly depend on the slip length. On the free-slip surface the droplet spreads out roughly twice as far as on the no-slip surface, and it takes about twice as long ($\tilde \tau_m \approx 2.0$ for no-slip and $\tilde \tau_m \approx 3.5$ for free-slip). For higher Weber numbers this discrepancy between no-slip and free-slip impacts further increases, but the qualitative mode of deformation remains the same.

To summarize: For $\text{We} < 1$ an impacting droplet deforms only slightly from spherical.  Then between roughly $\text{We} = 1$ and $\text{We} = 30$ it undergoes a transition from being puddle-shaped (thick rim compared to $D_0$) to being pizza-shaped (thin rim compared to $D_0$). In this transitional regime the flow can be very complex and is therefore difficult to characterize in general. For $\text{We} > 30$ the pizza shape becomes progressively more pronounced and no new transitions (in shape) occur.

From the above considerations the following message can be taken: \textit{if} some simple, general laws could be found for droplet impact, then one would expect them, at the very least, to be different in the regimes $\text{We}<1$ (spherical shape) and $\text{We}>30$ (pizza shape). At low Weber numbers, when the deformations are small, simple laws can be derived in a relatively straightforward manner \citep{Richard2000,Okumura2003}. For $\text{We} > 30$, on the other hand, it is not so clear whether this is possible (even in principle), as we will outline in the next section.

\section{Energy loss mechanisms\label{sec:lossmech}}

As mentioned in the introduction, the most straightforward approach to estimate $\tilde D_m$ is to assume that all of the initial energy $E_0 = E_{k0} + E_{s0}$ is used up by the work $E_s$ done against surface tension, with $E_{k0} = (\pi/12) \rho D_0^3 V_0^2$ and $E_{s0} = \pi\gamma D_0^2$ the kinetic and surface energy of the droplet before impact. If we exploit the fact that for $\text{We}>30$ the droplet spreads out into a thin disk (see section \S \sr{defmode}), then $E_s$ can be estimated as $E_s \approx (\pi/4)\gamma D_m^2 (1-\cos{\theta})$, where the term proportional to $(\cos\theta)$ accounts for the work done in expanding the wetted area with a dynamic contact angle $\theta$\footnote{For simplicity we assume that the contact angle remains constant during the spreading, which, in general, does not have to be the case, but it holds for our simulations.} \citep{Sikalo2005}. Equating $E_0$ and $E_s$ leads to
\eq{
	\tilde D_m(\text{We}, \theta) = \sqrt{\frac{4}{1-\cos\theta}\left[\frac{1}{12}\text{We} + 1\right]}.\label{eq:naive} 
}

As shown in figure~\fr{wrong}, this naive expression correctly captures the large Weber number trend $\tilde D_m \sim \sqrt{\text{We}}$ for free-slip impacts. However, the predicted pre-factor is completely off. For $\text{We} = 30$ and $\theta = 180^\circ$ equation \er{naive} predicts a spreading ratio of $\tilde D_m \approx 2.6$. This is about $30\%$ higher than the ratio found in our simulations (and in experiments). A first hint as to where this difference comes from was already discussed in the context of figure \fr{regimes}. The fact that some of the initial kinetic energy shows up as vortical motions in the rim, invalidates the assumption underlying equation \er{naive} that all of it is spend on surface energy \citep{Clanet2004}. As vortical motions are absent for $\text{We} = 300$, one might expect the above approximation to become better here, at least for the free-slip case, where there is no boundary layer. But as it turns out, it becomes even worse. Equation \er{naive} predicts $\tilde D_m \approx 7.2$, while the actual spreading ratio is $\tilde D_m \approx 5.1$, a difference of $41\%$.

\begin{figure}
\centerline{\includegraphics{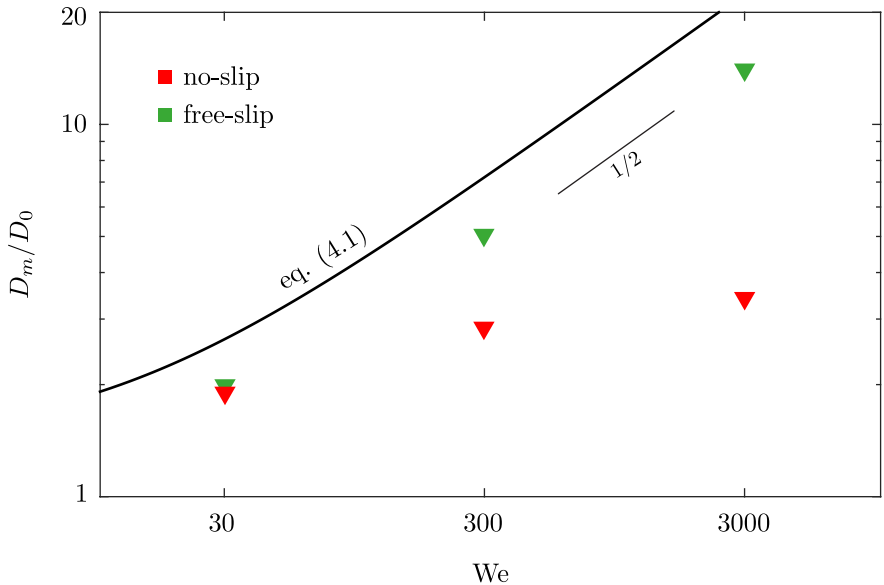}}
\caption{Maximum spreading ratio as a function of the Weber number, obtained from simulations with $\text{Re} = 500, \theta = 180^\circ$. The solid line shows the prediction from a naive energy balance between the initial kinetic energy and the final surface energy (equation \er{naive}).  \label{fig:wrong}}
\end{figure}

Since at high Weber numbers the ``left-over'' kinetic energy alone cannot account for the observed deviation from equation~\er{naive}, we must seek for other loss channels. In our simulations the only other possibility is viscous dissipation. Figures \fr{spatial}(a) and (b) show simulation snapshots for $\text{We} = 300$ (free-slip on the left, no-slip on the right) in which the red shading indicates the local dissipation rate $\epsilon_d$ in cylindrical~coordinates~($r$,~$z$):
\eq{
	\epsilon_d(r,z,t) &= 2\mu \left[ \left(\pd{v_r}{r}\right)^2 + \frac{v_r^2}{r^2} + \left(\pd{v_z}{z}\right)^2 \right] + \mu \left[\pd{v_r}{z}+\pd{v_z}{r}\right]^2, \label{eq:disip}
}
where $v_r$ and $v_z$ are the radial and vertical velocity, respectively. 

\begin{figure}
\centerline{\includegraphics{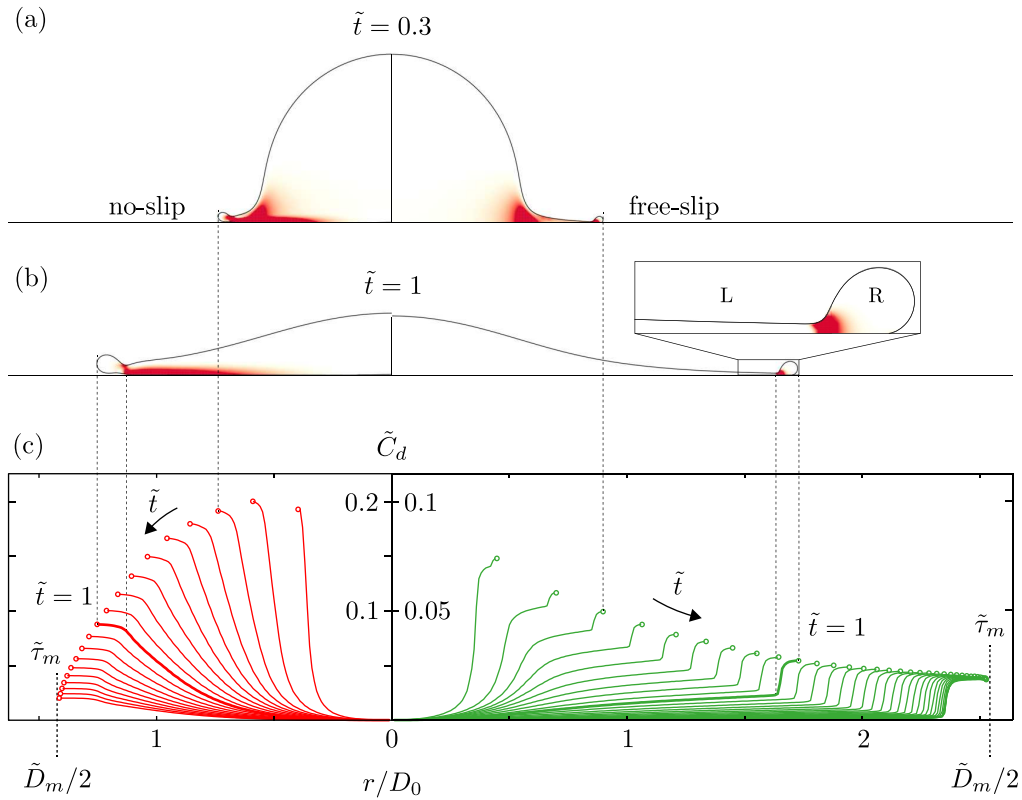}}
\caption{Visualization of viscous dissipation inside an impacting droplet ($\text{We} = 300$, $\text{Re} = 500$, $\theta = 180^\circ$). No-slip on the left, and free-slip on the right. (a-b)~Simulation snapshots at (a) ${\tilde t} = 0.3$ and (b) ${\tilde t} = 1$, in which the local dissipation rate $\epsilon_d$ is indicated by a red shading. For clarity, the intensity was clipped at $\epsilon_d = 10 \,\rho V_0^3/D_0$. For the free-slip droplet a close-up of the transition between the lamella (L) and rim (R) region is shown. (c)~Spatio-temporal evolution of the energy dissipation. The solid lines represent, for different times ${\tilde t}=0.1,0.2,...,{\tilde \tau}_m$, the cumulative dissipation rate, eq. \er{cumdis}, normalized by $\rho D_0^2V_0^3$. For each solid curve the final value corresponds to the total dissipation rate $\dot E_d$ at time $\tilde t$, i.e. $\dot E_d(\tilde t) = C_d(D_m/2,\tilde t)$. Mind the different scale on the y-axis for the two cases. The vertical dotted lines relate the features seen in figures (a-b) and~(c). \label{fig:spatial}}
\end{figure}

In the no-slip case the term $\mu (\pdi{v_r}{z})^2$ causes a significant amount of viscous dissipation near the solid surface, which somewhat occludes the picture. In the free-slip droplet this boundary layer is absent, and we can now easily identify other regions of high dissipation. A strong peak appears for example where liquid from the lamella enters the rim (see inset figure~\fr{spatial}(b)). This is reminiscent of the dissipative behaviour described in the introduction for a retracting rim (the Taylor-Culick problem). However, notice that in the initial stage of the impact (figure~\fr{spatial}(a)) there is also a large contribution to the dissipation where liquid from the central part of the droplet flows around the corner into the lamella. 

To see how these two different contributions evolve during the spreading, and how much they contribute to the total dissipation rate, $\dot E_d(\tilde t)$, it is convenient to look at cumulative plots of the dissipation rate, integrated along the $r$-axis, i.e.
\eq{
	C_d(r,\tilde t)=2\pi\int_0^r \int_0^{H_c}  r' \epsilon_d(r',z',\tilde t)\,dz' dr'.\label{eq:cumdis}
}
The lines $C_d(r)$ are shown in figure \fr{spatial}(c) for different times $\tilde t = 0.1, 0.2, ..., \tilde \tau_m$, and for $r$ running from the centre of the drop, $r=0$, to the tip of the rim at $r=R(\tilde t)$.  The final value, $C_d(R(\tilde t),\tilde t)$, of each of these lines then corresponds to $\dot E_d(\tilde t)$.

In the free-slip case two dissipative phases can be distinguished. For dimensionless times ${\tilde t} <1$ most of the dissipation happens where the liquid flows into the lamella, while for ${\tilde t}>1$ the peak near the rim becomes dominant (see figure \fr{spatial}(c), right panel). While the first contribution slowly diminishes in time, the contribution from the peak first increases and then approaches a constant value. For the impact parameters considered here it turns out that both phases contribute about the same amount to the total dissipation: $E_d = \int_0^{\tau_m} dt\; \dot E_d(t)$ (see also figure \fr{bars}).

During the no-slip impact the total dissipation rate $\dot E_d$ is initially roughly twice as high as that for the free-slip impact (mind the different scales on the axes). As the droplet spreads out, this high rate then quickly drops to about the same final value as observed for the free-slip case.  Dissipation in the shear boundary layer now overwhelms the two contributions described above, although some signs of those can still be seen in the snapshots (figures \fr{spatial}(a) and (b), left panels).


To conclude, during droplet impacts at $\text{We} > 30$ there can be at least \emph{four} different loss mechanisms at work: (i) left-over kinetic energy (in the form of vortical motions), (ii) viscous dissipation in the flow from the bulk into the lamella, (iii) viscous dissipation in the flow from the lamella into the rim, and, on a no-slip surface (iv) dissipation in the shear boundary layer. Only their \emph{combined} effect can explain the overall mismatch between equation \er{naive} and observations (simulations). Coming back to the concluding statement of the previous section, it may seem unlikely, at first sight, that this complex behaviour could give rise to a single simple law.

\section{Overall energy budget \label{sec:budget}}

Despite the pessimistic forecasts concluding the previous sections, a surprisingly simple picture emerges when we look at the \emph{final} energy distributions shown in figure \fr{bars}. For free-slip impacts (figure \fr{bars}(b)), the part of the initial kinetic energy that finally makes it into surface energy is virtually independent of We and Re. For (at least) two orders of magnitude in Weber number (from 30 to 3000) the change in surface energy amounts to roughly 1/2 of the initial kinetic energy. At $\text{We} = 30$, the remaining 1/2 is distributed approximately equally among kinetic energy and viscous dissipation, while at the higher Weber numbers it is mainly dissipated. In other words, it seems that for $\text{We} > 30$ (and free-slip) the overall energy loss is independent of both the detailed loss mechanism and the overall impact parameters.

\begin{figure}
\centerline{\includegraphics{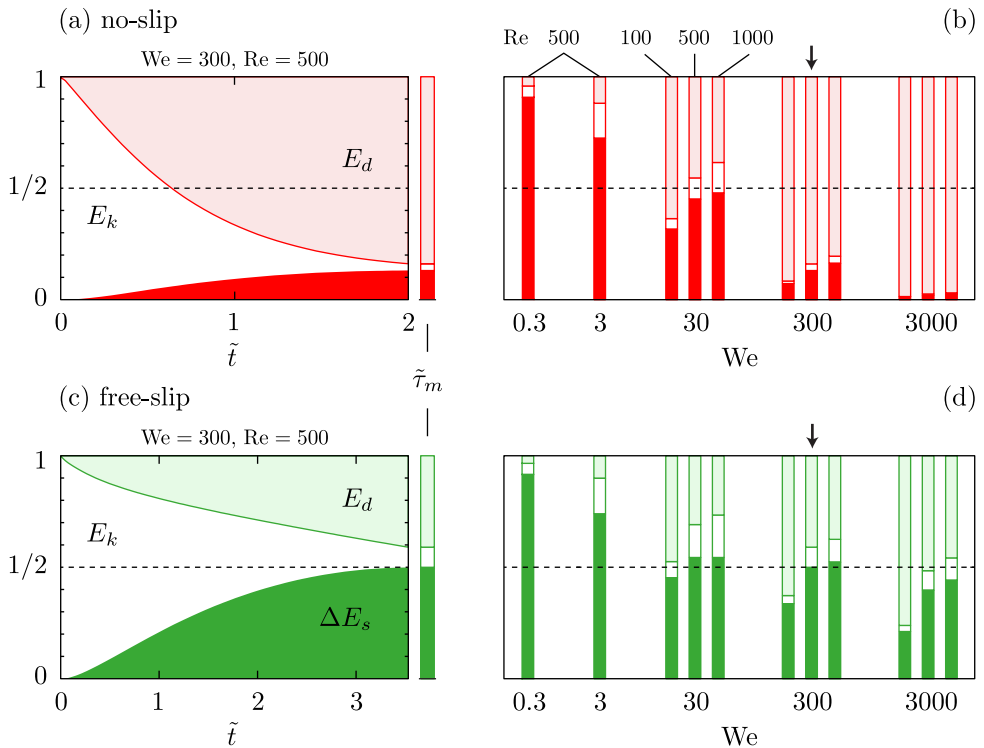}}
\caption{Energy budget for (a,b) no-slip impacts, and (c,d) free-slip impacts, normalized by the initial kinetic energy. The left graphs show for a reference case ($\text{We} = 300$, $\text{Re} = 500$, $\theta = 180^\circ$) how, over time, the kinetic energy $E_k$ is transformed into an increase $\Delta E_s$ of the surface energy and into heat $E_d$. The bar plots on the right show the final energy distribution for all simulations with $\theta = 180^\circ$ at the time of maximum spreading ${\tilde \tau}_m$. The dashed line indicates the 1/2 level. The arrows in (b) and (d) mark the cases for which the full time evolution is shown in (a) and (c). \label{fig:bars}}
\end{figure}

As outlined in the introduction, we believe that this `1/2-rule' is a manifestation of a geometrical head loss, that also occurs, for example, in suddenly expanding pipe flows and for retracting rims. In other words, the amount of dissipation seems to be imposed by the deformation mode of the droplet. Although we did not succeed to find a general proof of this rule for the highly transient flow situation presented by droplet impact (as can be done for steady pipe flow and rim retraction), experiment will be our judge (section \S\sr{experiment}).

As expected, the 1/2-rule breaks down for no-slip impacts (figure \fr{bars}(a)), for which dissipation in the boundary layer often also plays an important role (\S\sr{lossmech}), and for $\text{We} < 30$, where the mode of deformation drastically changes (\S\sr{defmode}). In figure \fr{bars}(b) we can also observe a deviation from the 1/2-rule when the Reynolds number gets too low compared to the Weber number, as is most clearly seen for the free-slip droplet with $\text{We}=3000$ and $\text{Re} = 100$. This marks the transition to a regime in which also viscosity greatly influences the deformation of the droplet. In the rim retraction problem this transition sets in for $\text{Oh}_h \equiv \mu/\sqrt{h\rho \gamma}  > 1$, where $\text{Oh}$ is the Ohnesorge number with as relevant length scale the thickness $h$ of the liquid film \citep{Sunderhauf2002, Savva2009}. If we tentatively assume that this result extends to the droplet impact case, with $D_0$ as length scale, i.e. $\text{Oh} \equiv \mu/\sqrt{D_0\rho \gamma} = \sqrt{\text{We}}/\text{Re}$, we find that for $\text{We}=3000$ a deviation from the 1/2-rule can be expected around $\text{Re} = \sqrt{3000} \approx 50$, consistent with the simulations. For $\text{We} < 1$ things become simple again as one here approaches the `elastic regime' of small deformations. In this regime practically \emph{all} of the initial kinetic energy is transformed into surface energy, and back \citep{Richard2000, Okumura2003, DeRuiter2014}. Assuming that the droplet deforms into a slightly oblate ellipsoid, one finds $D_m/D_0 - 1 = \sqrt{5/96}\sqrt{\text{We}} \approx 0.23 \sqrt{\text{We}}$ \citep{Richard2000}, which is consistent with experiments \citep{Okumura2003}.

The 1/2-rule for droplet impacts with $\text{We}>30$ and $\text{Oh} \ll 1$ is the main finding of this work. It can be concisely stated as follows: even if friction at the solid surface is negligible, still about one half of the initial kinetic energy is lost as far as its transformation to surface energy is concerned. On a no-slip surface it acts as a lower bound as $\text{Re} \rightarrow \infty$. 

In the next sections we will use this rule to derive simple analytical expressions for the spreading ratios in both free-slip and no-slip circumstances, and compare these to our simulation results and experiments.

\section{Analytical expressions for the spreading ratio in free-slip and no-slip impacts \label{sec:models}}

We will now return to the basic impact problem formulated in the introduction, and assess whether the discovered 1/2-rule can be applied to estimate the maximum spreading ratio $\tilde D_m = D_m/D_0$. This will allow for a direct comparison to a large set of experimental data.

For free-slip, the 1/2-rule states that the initial surface energy $E_{s0} = \pi\gamma D_0^2$ and about \emph{half} of the initial kinetic energy $E_{k0}/2 = (\pi/24) \rho D_0^3 V_0^2$ are transformed into the final surface energy $E_{s} = (\pi/4)\gamma D_m^2 (1-\cos{\theta})$. Equating these energies, $E_{s0} + E_{k0}/2 = E_s$, and solving for $D_m/D_0$, leads to:
\eq{
	\tilde D_m &= \sqrt{ \frac{4}{1-\cos{\theta}}\left(\frac{1}{24}\text{We} + 1 \right)} \qquad \text{(free-slip, $\text{We}>30$)} \label{eq:dmaxfs}
}

Qualitatively, this expression is similar to equation \er{naive}, but it should now also be \emph{quantitatively} correct, as we have properly accounted for the head loss $E_{d}^{\text{H}} \approx E_{k0}/2$. In figure~\fr{dmvswe} we have plotted equation~\er{dmaxfs} for $\theta = 90^\circ$ and $\theta = 180^\circ$, together with the maximum spreading diameters found in our simulations for these contact angles. For the free-slip cases with $\text{Re}=500$ and $\text{Re}=1000$ we find an excellent agreement. Of course this agreement was to be expected for the $180^\circ$ simulations, as these gave us the idea for the 1/2-rule in the first place. That it also works for $\theta = 90^\circ$ is evidence of its generality. In the next section a direct comparison to experiments, for which the Reynolds number was higher by an order of magnitude, will further strengthen this idea.

\begin{figure}
\centerline{\includegraphics{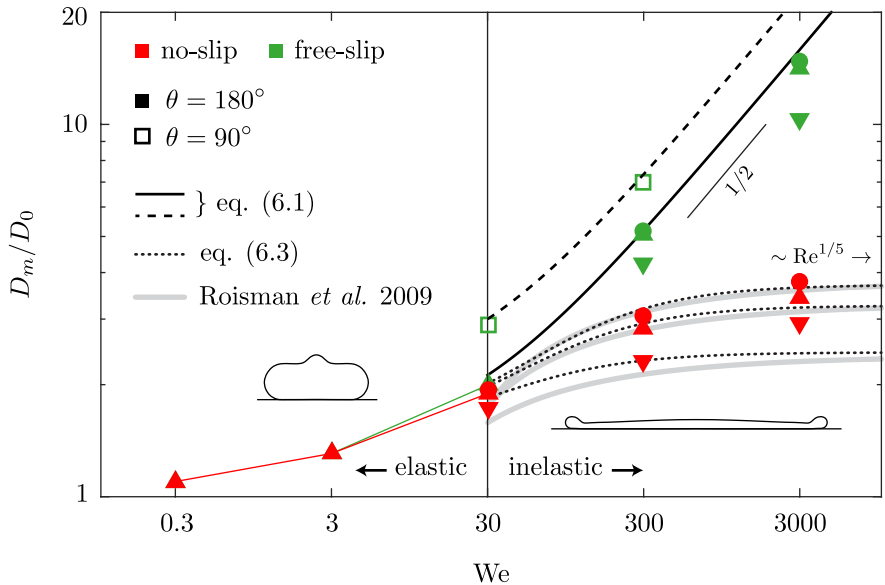}}
\caption{Maximum spreading ratios as a function of the impact Weber number found in simulations ran for Reynolds numbers of 100 ($\blacktriangledown$), 500  ($\blacktriangle$, {$\square$}) and 1000~({\large$\bullet$}), and contact angles of $\theta = 180^\circ$ ($\blacktriangledown$, $\blacktriangle$, {\large$\bullet$}), and $90^\circ$ ({$\square$}). Green symbols are used for impacts on a free-slip surface, and red symbols for impacts on a no-slip surface. The black curves show the predictions by equation \er{dmaxfs} (solid curve for $\theta=180^\circ$, dashed curve for $\theta = 90^\circ$) and equation \er{full} (dotted, $\theta = 180^\circ$). The insets show the typical shape of the droplets at low (puddle-shaped), and high Weber numbers (pizza-shaped). The vertical line at $\text{We}=30$ indicates the starting point of the fully developed inelastic regime (where the 1/2-rule holds). \label{fig:dmvswe}}
\end{figure}

As anticipated, the no-slip simulations show a large deviation from equation~\er{dmaxfs} at high Weber numbers. Here the spreading is also hindered by dissipation in the shear boundary layer at the solid surface. Integrating equation \er{disip} over both volume and time gives that this contribution scales as $E_d^{\text{BL}} \sim \mu (V_0/H_b)^2 (D_m^2 H_b) \tau_m$. For the thickness of the boundary layer we use, as before, $\tilde H_b \sim \sqrt{\tilde \tau_m / \text{Re}}$ (see e.g. \citet{Eggers2010}). The simplest (general) assumption we can make for the spreading time $\tau_m$ is that it is proportional to the maximum spreading diameter $D_m$ over the typical flow velocity $V_0$, i.e. $\tilde \tau_m \sim \tilde D_m$. It turns out that for $\text{We} > 30$ the relation $\tilde \tau_m =  (\tilde D_m-1)$ describes our simulation data quite well over a wide range of impact conditions (see figure \fr{tmvsdm}). Using this in our estimate of $E_d^{\text{BL}}$ we find
\eq{
	\frac{E_d^{\text{BL}}}{E_{k0}} \approx \alpha \frac{\tilde D_m^2\sqrt{\tilde D_m-1}}{\sqrt{\text{Re}}} \label{eq:bldis},
}
where $\alpha$ will be a fitting parameter of $\mathcal{O}(1)$.
In the limit where the retraction by surface tension is negligible ($\text{We} \rightarrow \infty$) and the spreading is large $\tilde D_m \gg 1$ we will have $E_d^{\text{BL}}/E_{k0} \sim 1$, so that equation \er{bldis} gives $\tilde D_m \sim \text{Re}^{1/5}$. This famous scaling law for the viscosity dominated regime, was previously derived from slightly different (but consistent) starting points \citep{Clanet2004, Roisman2009-2, Eggers2010}. The advantage of the derivation given here is that the underlying assumptions, and thereby equation~\er{bldis}, are still valid for finite Weber numbers, when the spreading is also hindered by the simultaneous retraction of the rim. If we assume, as a first approximation, that the dissipation in the boundary layer $E_d^{\text{BL}}$ and the head loss $E_{d}^{\text{H}}$ can simply be added to find the total dissipation $E_d$, then the energy balance $E_{k0} + E_{s0} = E_s + E_d$ in terms of $\tilde D_m$ becomes:
\eq{
	\frac{3(1-\cos\theta)}{\text{We}} \tilde D_m^2 + \frac{\alpha}{\sqrt{\text{Re}}}\tilde D_m^2 \sqrt{\tilde D_m -1} &= \frac{12}{\text{We}}+\frac{1}{2}.\qquad \text{(no-slip, $\text{We}>30$)}\label{eq:full}
}
One can easily check that this expression reduces to equation \er{dmaxfs} in the limit $\text{Re} \rightarrow \infty$. Although equation \er{full} does not allow for an explicit analytical solution for $\tilde D_m(\text{We},\text{Re}, \theta)$ it can be easily solved numerically. As shown in figure \fr{dmvswe}, with $\alpha \approx 0.7$ it predicts the spreading ratios found in the no-slip simulations reasonably well. In section \S\sr{experiment} and appendix \sr{appA} it is shown that the same value of $\alpha$ also works well for previous experimental data, covering a wide range of impact conditions. The observed deviations can be partly understood from a careful inspection of equation \er{disip} for the local dissipation rate. In places where the boundary layer and expansion flows overlap, cross-terms proportional to $\mu(\pdi{v_r}{z})(\pdi{v_z}{r})$ give additional contributions to the dissipation in the boundary layer. Also the strong assumption that the hindrance of the flow in the boundary layer does not affect the head loss, is probably not precisely true. In hindsight it is thus quite surprising that equation \er{full} works so well. Interestingly, for ($\text{Re} = 100$, $\text{We} = 3000$) equation \er{full} severely underestimates the spreading. As it turns out (see figure \fr{hcent}(a)), the combination of this low Reynolds number with a high Weber number, makes that a thin sheet is ejected from the edge of the droplet during impact. This sheet is lifted from the surface, so that the no-slip condition has no effect on this part of the droplet, leading to the larger than expected spreading in this special case.

\begin{figure}
\centerline{\includegraphics{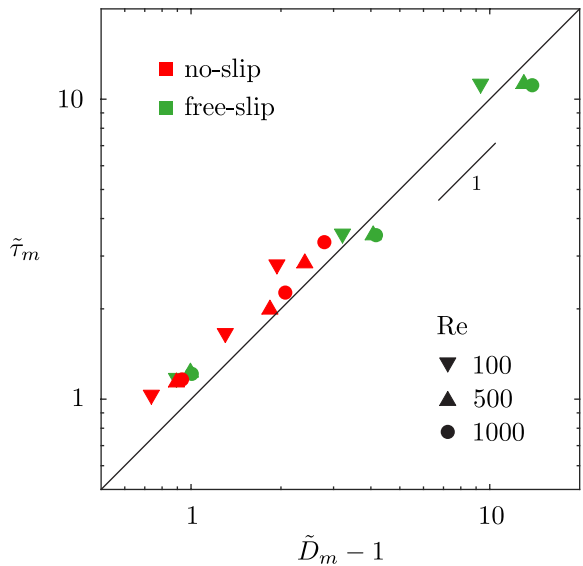}}
\caption{Spreading time $\tilde \tau_m$ versus spreading diameter $\tilde D_m$ for all simulations with $\text{We} > 30$. To better reveal the scaling, we plot $\tilde D_m-1$ on the horizontal axis and use a double logaritmic plot. The solid line indicates $\tilde \tau_m = \tilde D_m -1$. \label{fig:tmvsdm}}
\end{figure}

In the derivation of equation \er{full} we took the free-slip scenario as our basis and then corrected it for the dissipation in the shear boundary layer on no-slip surfaces. One can also approach this problem from the opposite side, and instead correct the large Weber number no-slip limit, $D_m \approx D_0\text{Re}^{1/5}$, for the simultaneous retraction of the rim \citep{Roisman2009}. Assuming that the rim retracts over a typical distance $L_c \approx V_c \tau_m$, where $V_c = \sqrt{2\gamma/h\rho}$ is the Taylor-Culick retraction velocity on a stationairy sheet, and using that in the shear dominated limit one has $\tilde \tau_m \sim \tilde D_m \sim \text{Re}^{1/5}$ and $\tilde h \sim \text{Re}^{-2/5}$ (see below) one arrives at the expression $\tilde D_m \approx a \text{Re}^{1/5} - b \text{We}^{-1/2}\text{Re}^{2/5}$, where $a$ and $b$ are adjustable parameters \citep{Roisman2009}. We found that $a = 0.95$ and $b = 0.70$ best describe our results (\cite{Roisman2009} originally proposed $a = 0.87$ and $b = 0.40$). As shown in figure 6, this expression captures our simulation data quite well. Likewise, it describes the recent results by \cite{Laan2014} as discussed in appendix \sr{appA}. However, we note that this expression does not approach the correct limit if the Reynolds number is send to infinity for a fixed Weber number, i.e. if the thickness of the boundary layer is small compared to the final thickness of the droplet. In this case it is more appropriate to use equation \er{full}, which (by design) does handle this limit correctly (see also appendix \sr{appA}).

\begin{figure}
\centerline{\includegraphics{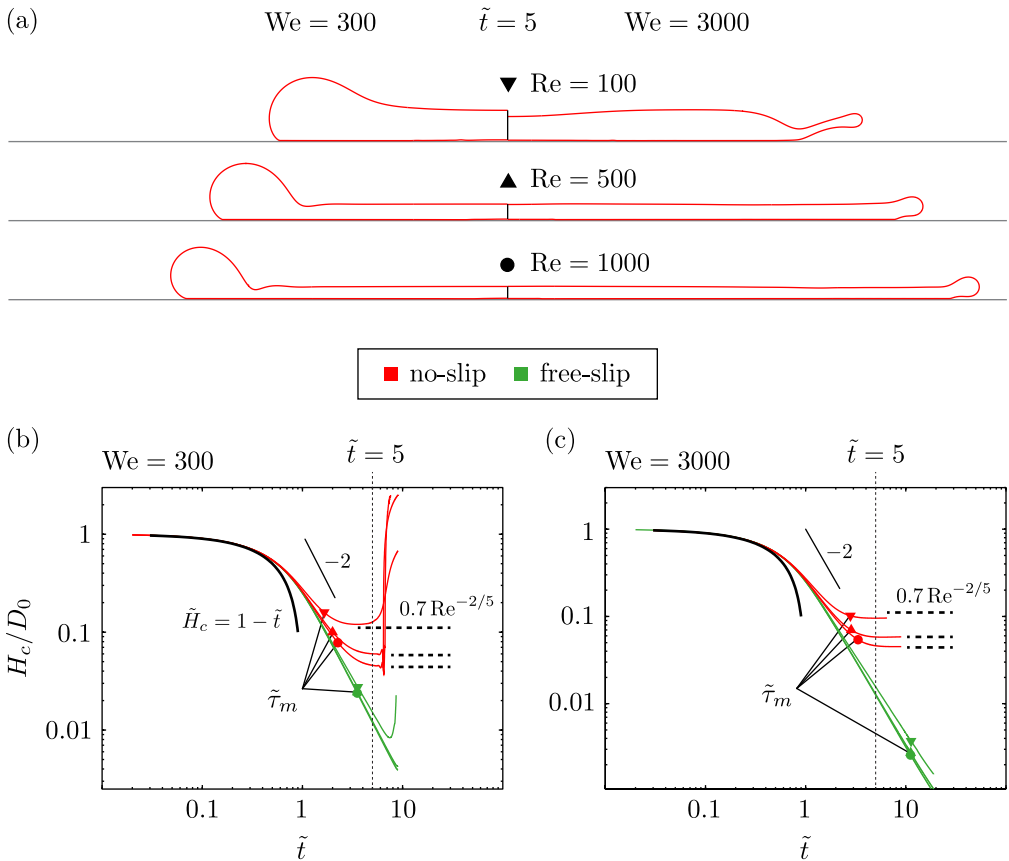}}
\caption{(a) No-slip droplet profiles at $\tilde t = 5$ for $\text{We} = 300$ (left) and $\text{We} = 3000$ (right) and different Reynolds numbers. For combinations of a low Reynolds number and a high Weber number, a thin sheet is ejected from the edge of the droplet, which is not hindered by the no-slip conditions at the solid surface. (b, c) Full time evolution of the central height $H_c(\tilde t)/D_0$ of the impacting droplets for (b) $\text{We} = 300$ and (c) $\text{We}=3000$. The black curves show how the height would change without impact. The logarithmic scale reveals an exponent of $-2$ during the spreading phase. For no-slip impacts (red) the height saturates to a plateau value of $0.7\:\text{Re}^{-2/5}$. The height of the free-slip droplets (green) keeps decreasing until it finally recoils. The times of maximum spreading $\tilde \tau_m$ are indicated by the thick symbols on each curve. \label{fig:hcent}}
\end{figure}

Although the maximum spreading ratio of no-slip droplets depends on \emph{both} We and Re, this is not the case for the minimum droplet thickness $H_m$ reached during the impact. As can been seen in figure \fr{hcent}(b), where we have plotted the time evolution of the central height $H_c(\tilde t)$, this minimum thickness is reached later than the maximum spreading and is well described by $\tilde H_m \approx 0.7 \text{Re}^{-2/5}$ for both $\text{We} = 300$ and $\text{We} = 3000$ (i.e. independent of We). As argued in \citet{Roisman2009,Eggers2010} the scaling can be understood as a collision of the upper part of the drop surface, which according to potential flow theory decreases in time as $\tilde H_c \sim 1/\tilde t^2$, and the boundary layer, which grows as $\tilde H_b \sim \sqrt{\tilde t/\text{Re}}$. In the absence of a thick rim (i.e. $\text{We} \rightarrow \infty$) the final droplet shape will approximate that of a disk. Using mass conservation, $\frac{2}{3} D_0^3 = D_m^2 H_m$, one recovers the upper bound $\tilde D_m \approx \text{Re}^{1/5}$ for no-slip impacts. For free-slip impacts, on the other hand, the central height of the droplet does not reach a minimum value and keeps on decreasing as $\tilde H_c \sim 1/\tilde t^2$ until rebound. These findings are in good agreement with recent experiments \citep{Lagubeau2012,Lastakowski2014}.


\section{Comparison to experiments \label{sec:experiment}}

Our simulations were done for a limited set of Reynolds numbers between 100 and 1000, but we hypothesised that the 1/2-rule and the equations derived from it,  \er{dmaxfs} and \er{full}, should hold quite generally. In figure \fr{expsim} we compare these models with available experimental data on (a) free-slip and (b) no-slip impacts. 

\begin{figure}
\centerline{\includegraphics{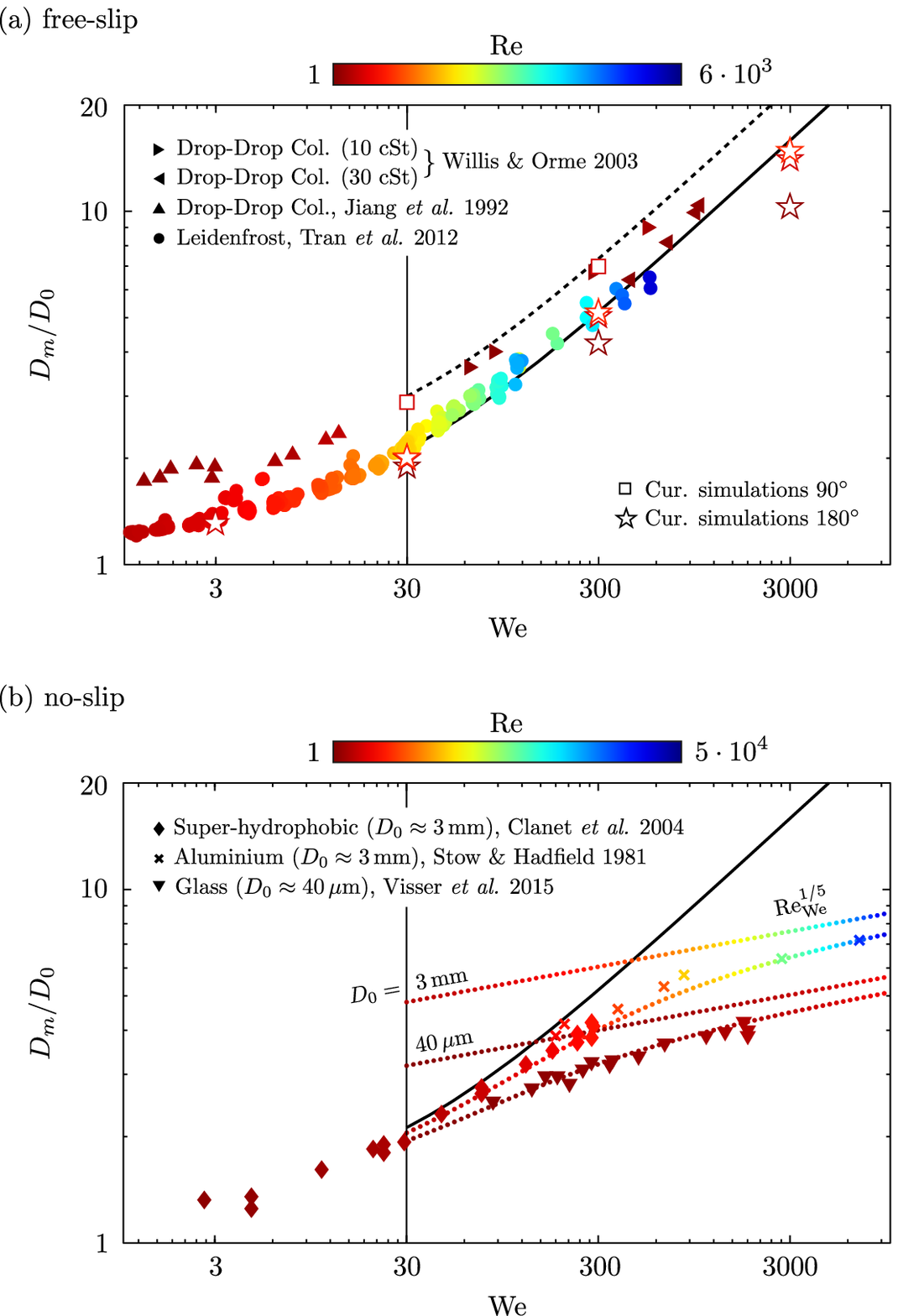}}
\caption{Comparison between the derived spreading models (lines) and (a) free-slip and (b) no-slip experiments (symbols). The colour of the symbols indicates the Reynolds number, which in experiments with a single liquid cannot be varied independently of the Weber number.  The black curves represent the model for free-slip impacts (equation \er{dmaxfs}, dashed line: $\theta = 90^\circ$, solid line: $\theta = 180^\circ$). In (a) the simulation results for free-slip are also shown (open symbols). In (b) the coloured dotted lines show the spreading limit $\text{Re}_\text{We}^{1/5} = \text{Oh}^{-1/5} \text{We}^{1/10}$ for two different droplet diameters ($D_0 = 3$~mm and $D_0 = 40$~$\mu$m) together with the full model, equation \er{full}. \label{fig:expsim}}
\end{figure}

Following \citet{Lastakowski2014} we expect the free-slip condition to hold for impacts of Leidenfrost droplets \citep{Tran2012}, which hover on a vapour film, and for symmetric droplet-droplet collisions \citep{Jiang1992, Willis2003}, where the mirror symmetry is equivalent to a free-slip condition and a contact angle of $90^\circ$~\citep{Roisman2009}. Figure \fr{expsim}(a) shows that there is a good agreement between the free-slip data and equation \er{dmaxfs}, especially if one takes into account that the Reynolds number reached in the experiments is one order of magnitude higher than that in our simulations. This provides strong evidence that the 1/2-rule is indeed independent of Re. Notice that the droplet collision data for the highest viscosity (30 cSt) lay somewhat below the $\theta = 90^\circ$ line, and come closer to the prediction for $\theta = 180^\circ$. This could either be a manifestation of the Reynolds number dependence we observed also in our simulations at low Reynolds numbers, or an indication that the droplets did not coalesce. However, the latter possibility is very unlikely, as these experiments were performed under vacuum conditions.

The no-slip condition generally holds for impacts on dry surfaces at room temperature, for which experimental spreading ratios obtained for water droplets \citep{Stow1981,Clanet2004, Visser2015} are shown in figure \fr{expsim}(b). The data points overlap with free-slip experiments at low Weber numbers, but start to deviate at larger Weber numbers. Here we expect them to bend towards the limit $\tilde D_m \approx \text{Re}^{1/5}$, as was the case in our simulations. However, whereas in the simulations we kept Re fixed as We was varied, in experiments it is usually the impact velocity $V_0$ that is varied, while the droplet diameter and liquid properties are kept fixed. For the free-slip prediction this is of no consequence, as it only depends on We, but for experiments on no-slip surfaces it means that the spreading limit effectively keeps shifting up as We is increased. For a given droplet diameter and liquid, we can predict this shift by expressing the Reynolds number in terms of We, as follows: $\text{Re}_\text{We} = (\rho\gamma D_0 \text{We})^{1/2}/\mu = \text{Oh}^{-1}\text{We}^{1/2}$, so that one expects the droplet diameters to stay below $\tilde D_m \approx \text{Re}_\text{We}^{1/5} = \text{Oh}^{-1/5} \text{We}^{1/10}$. In figure \fr{expsim}(b) we have plotted this relation for water droplets of $D_0 = 3$\,mm and $D_0 = 40\,\mu$m, corresponding to the liquid and diameters used in the experiments (straight dotted lines). The bended dotted lines represent equation \er{full} with $\theta = 180^\circ$ and $\alpha = 0.7$ as before. For both diameters this model fits the data well within experimental error. The match even seems to be somewhat better than for our simulations. This may indicate that the assumption that the dissipation in the boundary layer and the head loss can be added independently, is satisfied more closely for these experimental parameters. The use of a constant dynamic contact angle of 180$^\circ$ in the model is justified by the fact that for high Weber number impacts, wettability effects play only a minor role compared to inertia and viscous shear \citep{Rioboo2002}. Furthermore, in experiments it is seen that the dynamic contact angle of an impacting drop is generally closer to $180^\circ$ than to their static value \citep{Sikalo2005}.

\section{Conclusion \& outlook}

Inelastic droplet impact ($\text{We} > 30$) seems to fall into a special class of fluid problems in which the total energy loss associated with surface deformations does not explicitly depend on the properties of the liquid, nor on the detailed loss mechanism(s). This universal head loss is significant, as it always amounts to about 1/2 of the initial kinetic energy, even when all other sources of dissipation are negligible. 

When this 1/2-rule is incorporated in a simple energy balance one obtains an accurate analytical expression for the spreading ratio on free-slip surfaces  (equation \er{dmaxfs}). For impacts on no-slip surfaces this result must be corrected for dissipation in the shear boundary layer. The resulting expression (equation \er{full}) is in reasonable agreement with both simulations and experiments and is consistent with previous modelling approaches. The advantage of an energy balance approach is that it can generally be easily adapted to new scenarios. One can, for example, include a finite slip length $\lambda$ into the model by simply replacing the estimate $\epsilon \sim \mu (V_0/H_b)^2$ by  $\epsilon \sim \mu (V_0/(H_b+\lambda))^2$ in the derivation of equation~\er{full}, or even incorporate non-Newtonian effects \citep{Boyer2016}.

The 1/2-rule breaks down for $\text{We} < 30$ where a different mode of deformations sets in as one approaches the elastic regime observed for $\text{We} < 1$. Coincidentally the behaviour for $\tilde X = \tilde D_m - 1$ in the elastic regime, $\tilde X \approx 0.23 \sqrt{\text{We}}$ \citep{Richard2000,Okumura2003}, is almost the same as that for $\tilde D_m$ for free-slip impacts at high Weber numbers (for $\theta = 180^\circ$), $\tilde D_m = \sqrt{\text{We}/12} \approx  0.29\sqrt{\text{We}}$. A plot of $\tilde D_m - 1$ versus We on a logarithmic scale will therefore give the impression of a uniform scaling for $\tilde D_m - 1$ over the whole range from elastic to inelastic impacts. However, this does not have any physical significance, as the origin of the numerical pre-factors is very different in both cases. 

Also when the Reynolds number is small compared to the Weber number, deviations from the 1/2-rule can be observed. Based on the analogy to the Taylor-Culick problem we speculated that this transition sets in for $\text{Oh} = \sqrt{\text{We}}/\text{Re} > 1$. This is consistent with our current simulations, but a more systematic study around this transition would have to be performed to put this idea on more firm grounds.  



We hope that this study provides a useful new view on the old problem of droplet impact. On the other hand, the universal head loss discovered for this complex transient flow problem raises the interesting question as to its general applicability to free-surface flows involving large deformations. It will be interesting to study the involved boundary layer dynamics and the coupling between flow and free-surface surface deformation on a more fundamental level in simpler flow geometries. With the current advent of reliable and flexible numerical fluid solvers like Gerris, such a systematic approach is well within reach.

\acknowledgements
We thank Rodolfo Ostilla M\'onico, Rianne de Jong, Marie-Jean Thoraval and Minori Shirota for a careful reading of the first manuscript and fruitful discussions. This project received funding from an ERC Advanced Grant and from the NWO Spinoza programme.

\appendix
\section{Comparison of no-slip droplet impact models to experiments \cite{Laan2014}\label{sec:appA}}

In the main text (\S\sr{models}) we discussed two complementary approaches to model the impact of a droplet onto a no-slip surface. One approach is to take the free-slip case as a basis and correct it for dissipation in the shear boundary layer (equation \er{full}). The other approach is to start from the limit in which the thickness of the spreading lamella is set by the boundary layer, so that $\tilde D_m \approx a \text{Re}^{1/5}$, and then subtract a correction, $b \text{Re}^{2/5}\text{We}^{-1/2}$, for the simultaneous retraction of the rim on this film \citep{Roisman2009}. We found that both approaches capture our no-slip simulation data reasonably well. Here we further test the models and the parameters proposed for them ($\alpha = 0.70$ and $a = 0.95, b=0.70$, respectively) by a direct comparison to an extensive set of experimental data kindly provided to us by \cite{Laan2014}. 

As shown in figure \fr{bonn}, overall both models agree well with the experimental data. As can be expected the model by \cite{Roisman2009} (figure \fr{bonn}c) is somewhat more accurate for the larger spreading ratios, where the shear boundary layer dominates the dynamics, but it shows an underestimation (especially for the low viscosity water) for $\tilde D_m < 2.5$. Here the boundary layer (with thickness $H_b$) does not occupy the whole thickness of the spreading lamella. As a consequence the retraction velocity (based on $H_b$ in this model) is overestimated (and the spreading is thus underestimated). Equation~\er{full} on the other hand does not assume that the lamella thickness is set by the boundary layer and it therefore does not have this problem (figure \fr{bonn}d). However, for this model the match is somewhat less good at larger spreading ratios.

\begin{figure}
\centerline{\includegraphics{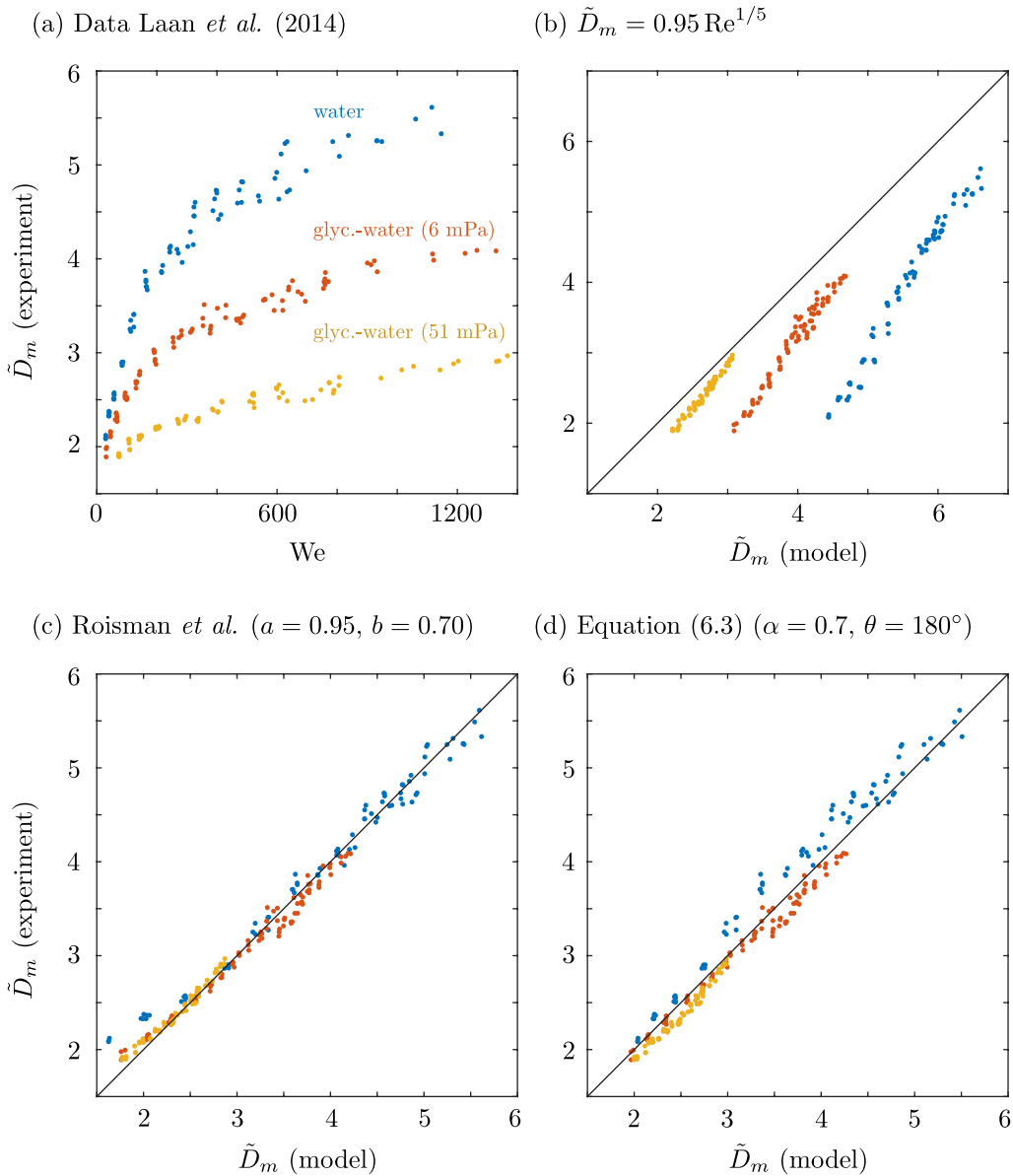}}
\caption{Comparison of no-slip droplet impact models to the experimental data in \cite{Laan2014}. (a) Spreading data as function of Weber number for three different water-glycerol mixtures. (b) Basic model in which the droplet spreads into a disk with the thickness set only by the shear boundary layer. (c) Modified model proposed by \cite{Roisman2009} in which the (estimated) distance over which the rim retracts is explicitly subtracted. (d) Current model (equation \er{full}) in which the free-slip scenario is corrected for dissipation in the shear boundary layer.  \label{fig:bonn}}
\end{figure}

\clearpage

\bibliography{refs}
\bibliographystyle{jfm}

\end{document}